\documentclass[aps,prd,showpacs,eqsecnum,twocolumn,superscriptaddress]{revtex4}

\usepackage{latexsym} \usepackage{amssymb} \usepackage{amsfonts}
\usepackage{amsmath} \usepackage{bm} \usepackage[dvips]{graphicx}
\usepackage{color}
\usepackage{subfigure} \usepackage{times} \usepackage{units}
\usepackage{hyperref}

\usepackage[utf8x]{inputenc} \usepackage{amssymb,amsmath}
\usepackage{graphicx} \usepackage[squaren]{SIunits} 

\begin{document}
\title{Denoising of gravitational-wave signal GW150914 via total-variation methods}
\author{Alejandro \surname{Torres-Forn\'e}}\affiliation{Departamento de
  Astronom\'{\i}a y Astrof\'{\i}sica, Universitat de Val\`encia,
  Dr. Moliner 50, 46100, Burjassot (Val\`encia), Spain} 

\author{Antonio \surname{Marquina}}\affiliation{Departamento de
  Matem\'atica Aplicada, Universitat de Val\`encia,
  Dr. Moliner 50, 46100, Burjassot (Val\`encia), Spain}  
  
\author{Jos\'e A. \surname{Font}}\affiliation{Departamento de
  Astronom\'{\i}a y Astrof\'{\i}sica, Universitat de Val\`encia,
  Dr. Moliner 50, 46100, Burjassot (Val\`encia), Spain}
  \affiliation{Observatori Astron\`omic, Universitat de Val\`encia, C/ Catedr\'atico 
  Jos\'e Beltr\'an 2, 46980, Paterna (Val\`encia), Spain}
  
\author{Jos\'e M. \surname{Ib\'a\~nez}}\affiliation{Departamento de
  Astronom\'{\i}a y Astrof\'{\i}sica, Universitat de Val\`encia,
  Dr. Moliner 50, 46100, Burjassot (Val\`encia), Spain}   
    \affiliation{Observatori Astron\`omic, Universitat de Val\`encia, C/ Catedr\'atico 
  Jos\'e Beltr\'an 2, 46980, Paterna (Val\`encia), Spain}


\begin{abstract}
We apply a regularized Rudin-Osher-Fatemi total variation (TV) method to denoise the transient gravitational 
wave signal GW150914. We have previously applied TV techniques to denoise numerically generated gravitational 
waves embedded in additive Gaussian noise, obtaining satisfactory results irrespective of the signal morphology 
or astrophysical origin. We find that the non-Gaussian, non-stationary noise from the gravitational wave event 
GW150914 can also be successfully removed with TV-denoising methods. The quality of the denoised 
waveform is comparable to that obtained with the Bayesian approach used in the discovery 
paper~\cite{GW150914-prl}. TV-denoising techniques may thus offer an additional viable approach for 
waveform reconstruction.
\end{abstract}

\pacs{
04.30.Tv,	
04.80.Nn,	
05.45.Tp,	
07.05.Kf,	
02.30.Xx.	
}
\maketitle

\section{Introduction}
\label{section:intro}

With the unprecedented detection of the transient gravitational-wave (GW) signal  GW150914 on September 14, 
2015 by the two Advanced-LIGO interferometers~\cite{GW150914-prl} GW astronomy has finally become a reality. 
The signal, that was detected at a peak GW strain of $10^{-21}$, matches consistently numerical relativity
waveforms~\cite{SXS, Mroue:2013, campanelli:2006} for the final few cycles (chirp) and merger (burst) of the 
coalescence of two stellar-origin black holes in a binary system, along with the subsequent ringdown
signal of the resulting final black hole. The statistical significance of the observation has been reported to be 
greater than $5.1\sigma$~\cite{GW150914-prl}.

For coalescing compact binary signals such as GW150914, the inspiral part of the signal can be detected by either 
targeting a broad range of generic transient signals or by correlating the data with analytic waveform templates from 
general relativity and maximizing such correlation with respect to the waveform parameters. When the signal-to-noise 
ratio (SNR) of the filter output over a wide bandwith of the detector exceeds an optimal threshold,  the matched-filtering 
technique generates a trigger associated with a specific template. For low SNR signals, the inherent non-stationarity 
and non-Gaussanity of the detector noise renders however the identification of the signal fairly challenging.
In the case of GW150914, with a SNR of 24, two independent waveform reconstructions were applied in~\cite{GW150914-prl} namely a binary black hole template waveform from~\cite{abbott-1} and a Bayesian approach based on a linear combination 
of sine-Gaussian wavelets and a parameterized model of non-stationary, non-Gaussian noise~\cite{cornish:2015,abbott-2}. 
With the advent of {\it real} GW data for the first time, it is worth investigating how  approaches for GW reconstruction and 
denoising {\it alternative} to those used in~\cite{GW150914-prl} do actually perform. That is the aim of this paper. More 
precisely, we present in this work the results of applying Total Variation (TV) denoising techniques on the publicly 
available waveform data for GW150914 and compare our findings with the waveform reconstruction methods reported 
in~\cite{GW150914-prl}.

In~\cite{alex} we recently presented such TV-norm regularized methods for denoising and detection of GWs embedded 
in additive Gaussian noise. In our approach, a new regularization term is added to the error function (fidelity), weighted 
by a positive Lagrange multiplier which measures the relative importance of the data-dependent fidelity term. Key to this
new approach is the fact that the regularization strategy is based on a $\text{L}_1$-norm, whose main advantage is that 
it favors {\it sparse} solutions. In this paper we propose an iterative procedure to solve denoising problems based on the 
so-called regularized Rudin-Osher-Fatemi (rROF) algorithm (see~\cite{Rudin:1992,alex}). The rROF method 
runs a new scale space from finer to coarser scales, with a termination criterion given by a discrepancy principle.
Usually, this criterion is enforced  by matching the square of the $\text{L}_2$-norm of the residual with the variance 
of the noise, if the latter is known. Since this is not the case for GW data, our iterative procedure is terminated as soon 
as the reconstructed signal starts to loose amplitude around local extrema. As we show below, this procedure returns 
an accurate denoising of the signal  GW150914. The interested reader is addressed to~\cite{Goldstein:2009} and 
references therein for details and applications of TV-denoising methods in fields as diverse as medical imaging, radar 
imaging, and magnetic resonance imaging, and to our previous work~\cite{alex} for the first specific application of this 
method in numerically-simulated GW signals.

This paper is organized as follows: in Section~\ref{section:method} we briefly summarize the main 
aspects of our numerical procedure. Next, in Section~\ref{section:results} we present the results of applying 
our TV-denoising method to GW150914 and compare with the waveform reconstructions reported 
in~\cite{GW150914-prl}. Finally,  Section \ref{section:summary} presents the conclusions of our study.

\section{Review of the method}
\label{section:method}

We start by providing an overview of our TV method. Full details are presented in Ref.~\cite{alex}.
We shall assume the general linear degradation model 
\begin{equation}
\label{eq:denoising_problem}
f=u+n~,
\end{equation}
where $f$ is the observed signal, $n$ is the noise and $u$ is the signal to be recovered. Signal denoising
stands for estimating a noise-free signal $u$
whose square of the $L^2$-distance to 
the observed noisy signal $f$ 
is the variance of the noise, i.e. 
\begin{equation}\label{eq:fidelity_term}
||u-f||_{\text{L}_2}^2=\sigma^2~,
\end{equation}
where $\sigma$ denotes the standard deviation of the noise. It is well-known that $\text{L}_2$ norm (i.e.~least squares) models
to solve the denoising problem present deficiencies such as the appearance of Gibbs phenomena or non-unique
solutions. Those are overcome by regularizing the least squares problem using an auxiliary energy (`prior') 
$R(u)$, and solving a constrained variational problem
\begin{eqnarray}
\label{eq:constrained}
\underset{u} {\min} \, R(u) \hspace{0.3cm}
 \mbox{subject to} \hspace{0.3cm} ||f-u||^2_{L^2}=\sigma^2~,
\end{eqnarray}
where the functional $R(u)$ measures the quality of the signal $u$, namely~the smaller 
$R(u)$ the cleaner the signal.  If the energy $R(u)$ is convex, problem (\ref{eq:constrained}) has a 
unique solution. Interestingly, there exists an unconstrained version of the variational problem, obtained by 
adding the constraint (the ``fidelity term") weighted by an unknown, positive Lagrange multiplier $\mu>0$ 
to $R(u)$
\begin{equation}
 \label{eq:unconstrainL2}
 u=\underset{u} {\text{argmin}}\left\{R(u)+\frac{\mu}{2} \, ||f-u||_{\text{L}_2}^2\right\}~,
 \end{equation}
 a procedure known as Tikhonov regularization. There exists a unique value of $\mu>0$ such that the unique solution $u$ 
matches the constraint.

Rudin, Osher and Fatemi proposed in \cite{Rudin:1992} the TV norm as regularizing functional for the variational 
model for denoising (\ref{eq:unconstrainL2})
\begin{equation}\label{tv}
\mathrm{TV}(u)=\int |\nabla u| \,,
\end{equation}
where the integral is defined on the domain of the signal~\footnote{Note that this differs from the most common Wiener filter model for solving (\ref{eq:unconstrainL2}) in which $R(u)=\int{|\nabla u|^2}$. }. Therefore, the ROF model amounts to solving the following variational problem for denoising:
\begin{equation}
\label{eq:rof}
u=\underset{u} {\text{argmin}}\left\{\mathrm{TV}(u)+\frac{\mu}{2}||u-f||_{\text{L}_2}^2\right\}~.
\end{equation}
The TV norm energy is essentially the $\text{L}_1$-norm of the gradient of the signal. Thus, the ROF model reduces 
noise by {\it sparsifying} the gradient of the signal and avoiding spurious oscillations (ringing).

In this paper we use a regularized ROF algorithm 
by solving the associated Euler-Lagrange equation of an
energy that includes a smoothed TV norm (see~\cite{alex} for details).
The rROF algorithm is used as a building block of an iterative procedure that
runs the scale space from the original noisy signal to the processed signal.
Roughly speaking, we first choose the regularization parameter $\mu$ equal to a constant value $\mu_0$, 
which is larger than the optimal value needed to obtain a denoised 
signal by means of the application of the rROF algorithm. The value of $\mu_0$
is kept fixed through all the iterations. Next, we compute $u_1$ by solving
\begin{eqnarray}
u_1&=&\underset{u} {\text{argmin}}\left\{\mathrm{TV}(u)+\frac{\mu_0}{2}||u-f||_{\text{L}_2}^2\right\}~,
\\
f&=&u_1+v_1~,
\end{eqnarray}
where $v_1$ is the residual.
Then, we apply again the rROF algorithm using the same $\mu_0$ and taking
as input signal $u_1$ to obtain $u_2$. We thus have
\begin{eqnarray}
u_1&=&u_2+v_2~.
\end{eqnarray}
Applying this procedure for an arbitrary number of times $n$ we
obtain a sequence of signals $u_n$ for $n=1, \cdots$ such that
\begin{eqnarray}
u_{n-1}&=&u_n+v_n~,
\\
f&=&u_n + \sum_{i=1}^{n} v_i~.
\end{eqnarray}
The iteration stops when some discrepancy principle is satisfied,
i.e.~when the square of the $\text{L}_2$-norm of the residual matches the variance of the noise.
In practice, however, the variance of the noise is not available and we have to resort to some
other termination criterion. Therefore, the iterative procedure is stopped
at some $n_0$ which is selected to make it coincide with the appearance of the
denoised signal before its local extrema start loosing total variation. We regard signal $u_{n_0}$ 
as the denoised signal. Note that the scale space defined by this iteration is 
different from the one observed when the parameter
$\mu$ runs from small to larger values, as it is done in the Bregman refinement 
iterative algorithm (see~\cite{Bregman:1967,Osher:2005}).
The advantage of using the iterative procedure proposed here
is that local extrema and edges are significantly better preserved when the scale space 
is run in the opposite direction 
(towards smaller values of $\mu$). 

\section{Results}
\label{section:results}

\begin{figure*}[t]
	\begin{subfigure}
	{\includegraphics[width=81mm]{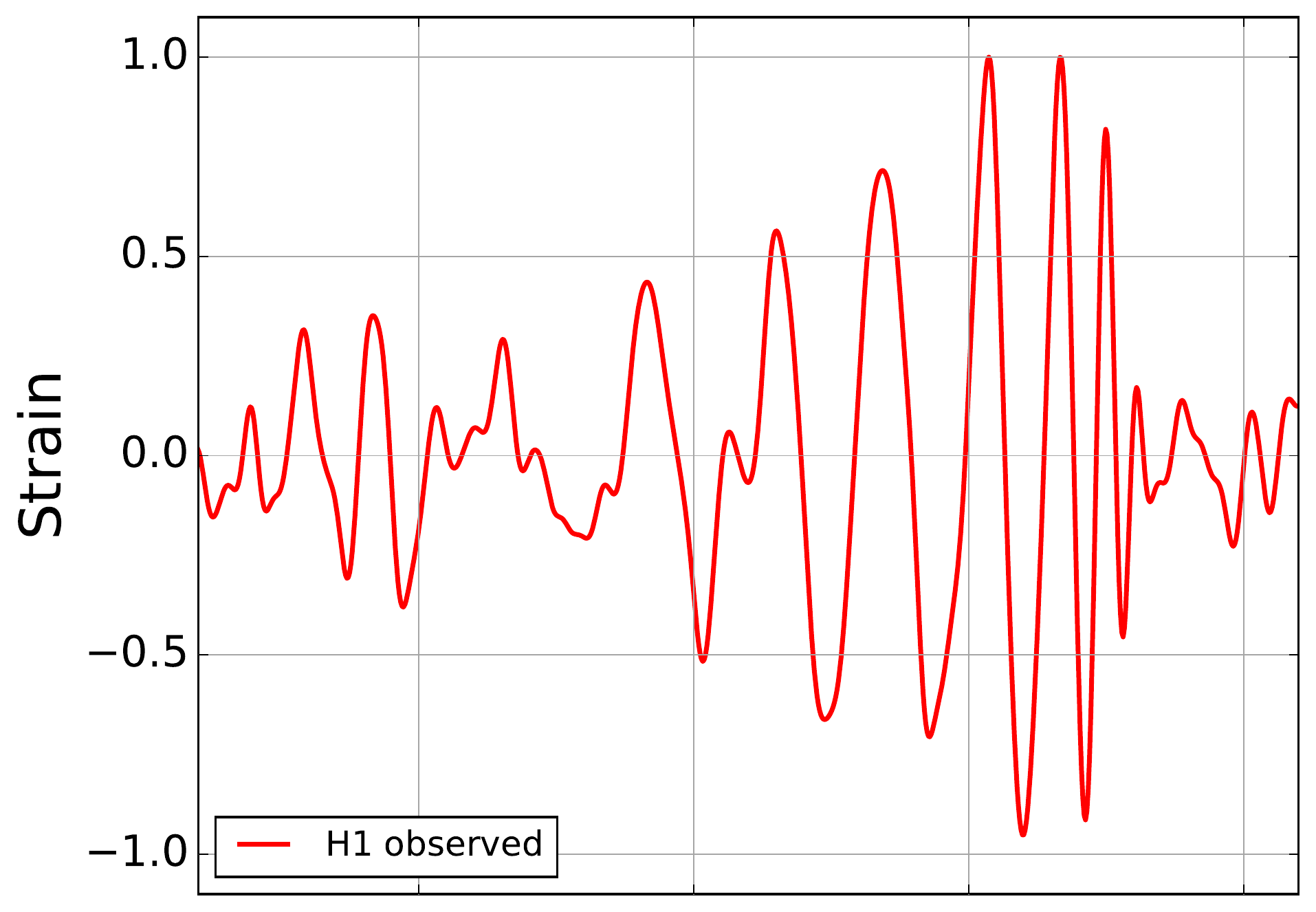}}
	\end{subfigure}
	\begin{subfigure}
	{\includegraphics[width=70mm]{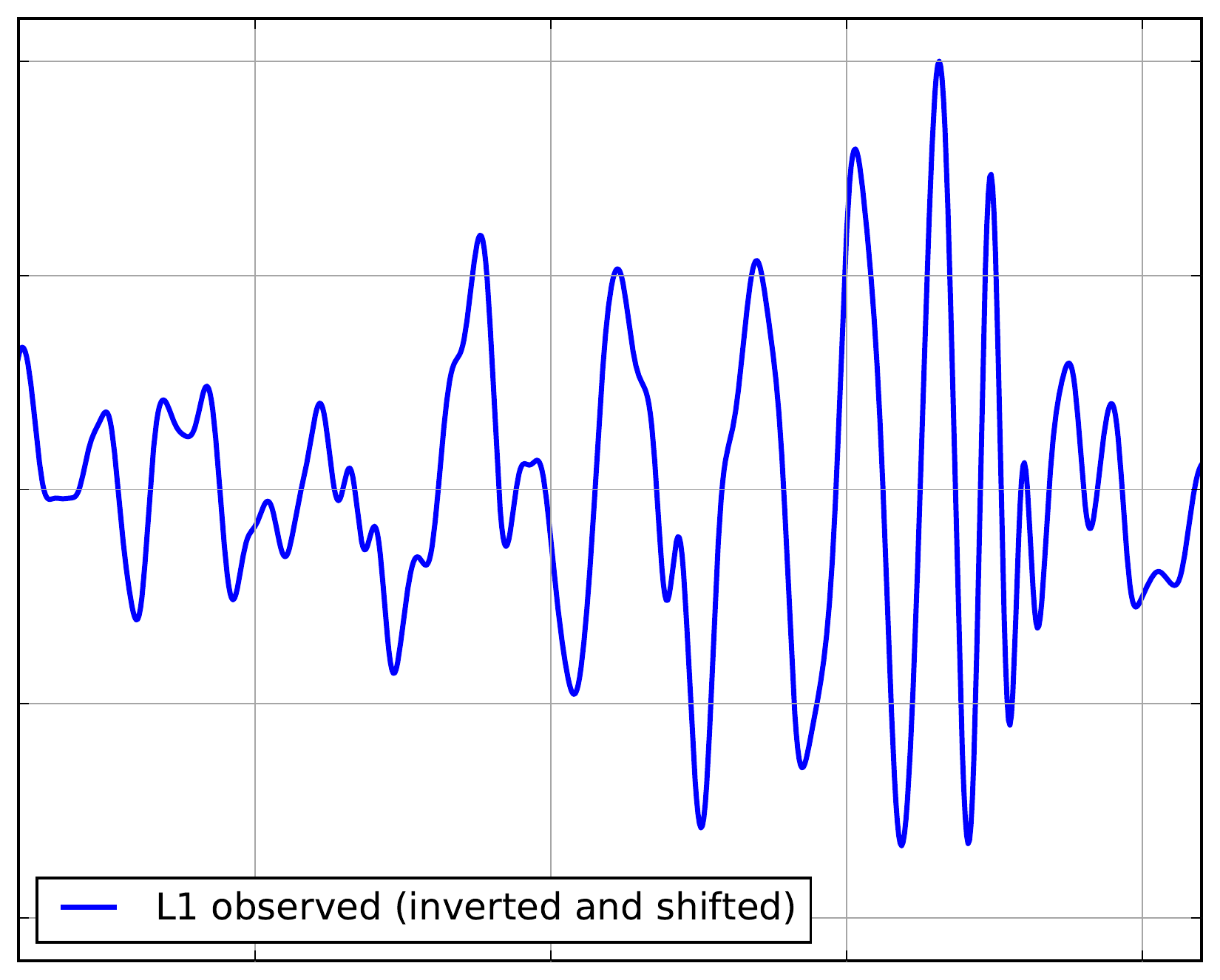}}
	\end{subfigure}\\ \vspace{-3mm}
	\begin{subfigure}
	{\includegraphics[width=81mm]{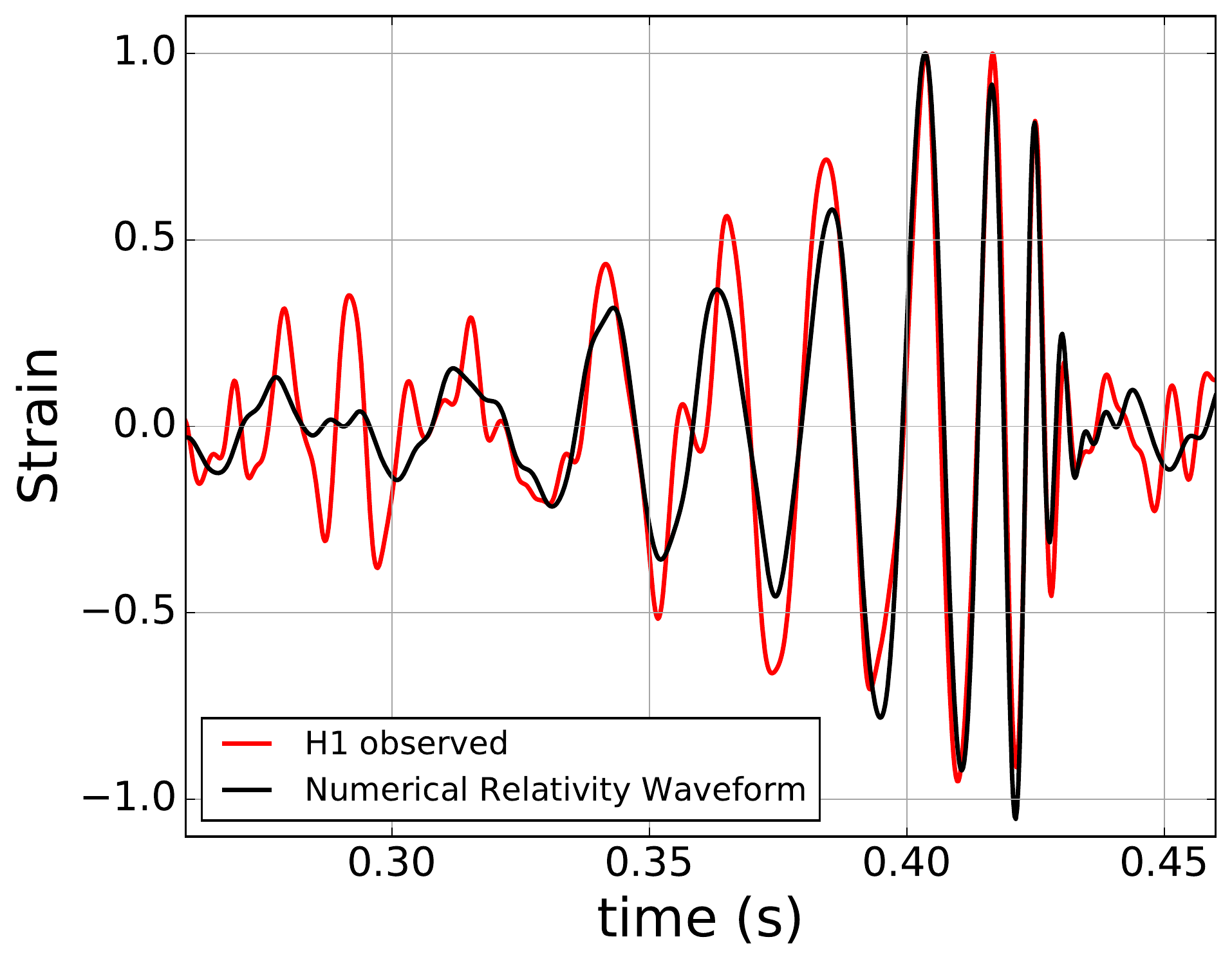}}
	\end{subfigure}
	\begin{subfigure}
	{\includegraphics[width=70mm]{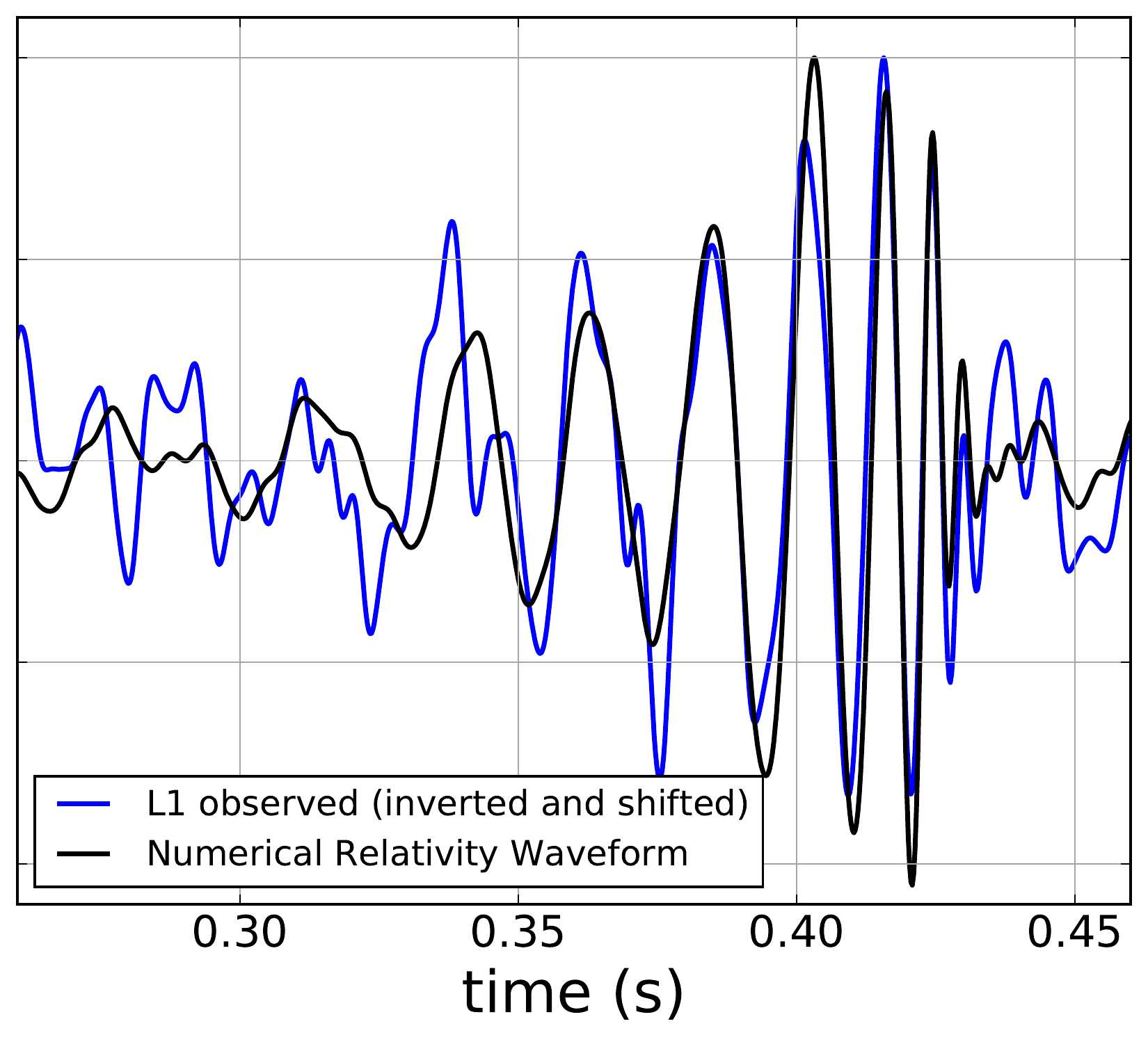}}
		\end{subfigure}
	\vspace{-1mm}
	\caption{Top panel: Results of applying the iterative rROF algorithm to the gravitational-wave event GW150914 observed by the LIGO Hanford (left plot) and Livingston (right plot) detectors. Bottom panel: Comparison of our denoised waveforms (red and blue colors) with the processed binary black hole template from numerical relativity~\cite{abbott-1} (black curve). 
The rROF algorithm has been used with $\mu_0=0.2$ and 10 iterations. The times shown in the $x$-axis are as in~\cite{GW150914-prl}. }
	\label{fig:Strain}
\end{figure*}

We turn to describe the results of applying the rROF method in the time domain to the 
strain time-series data associated with the gravitational wave event GW150914. These data 
have been released by the LIGO Open Science Center~\cite{GW150914data}.
Our method is transparent to the sampling frequency of the data. GW150914 data are sampled at
both 4 kHz and 16 kHz. We have found that our results are very similar for both sampling frequencies. Therefore,
for efficiency reasons, our analysis is performed using data corresponding to a sample frequency of 4 kHz.

To assess our denoising procedure we have tried to use as few assumptions about instrumental noise as possible, due to the fact that the detector noise is non-Gaussian and non-stationary. However, a minimum noise preprocessing is required due to two main reasons. On the one hand there are well-known, modeled sources of narrow-band noise, such as the electric power (at 60 Hz and higher harmonics), mirror suspension resonances or calibration lines (see Fig.~3 of~\cite{GW150914-prl}). On the other hand, ground-based detectors such as LIGO are not sensitive to low frequencies because of seismic noise. Therefore, we highpass the time series above 30 Hz to remove seismic noise and, following~\cite{GW150914-prl} we also filter out all spectral lines of both detectors.

Once the time-series has been preprocessed, we apply the rROF method in the time domain using the iterative
procedure described in the previous section. 
As we did in~\cite{alex}, to ensure the best conditions for the convergence of the algorithm and to avoid round-off
errors, the data are normalized to vary between -1 and 1 before the application of the algorithm. We choose a relatively 
high value of the regularization parameter $\mu_0$, larger than the optimal value (see~\cite{alex}), namely $\mu_0=0.2$.  Our analysis shows that less than 10 iterations are enough to denoise the signal and obtain an accurate and smooth result. We note that we can also apply the algorithm with a single iteration using the ``optimal" value of the regularization parameter, significantly lower than the selected $\mu_0$. As we showed in~\cite{alex}, the use of low values of $\mu$ leads to larger noise removal and more smooth-looking results (suppressing high-frequency noise in the denoised signals). However, we already noticed that for gravitational waveform templates of binary black holes embedded in Gaussian noise, low values of $\mu$ cannot capture properly the high frequency part of the signal, i.e.~the merger and the ringdown (see~\cite{alex} for details).  The iterative procedure we employ here for the first time allows us to remove noise in a more progressive way while capturing at the same time the high-frequency part of the signal.

\begin{figure*}[t]
    \begin{subfigure}
    {\includegraphics[width=90mm]{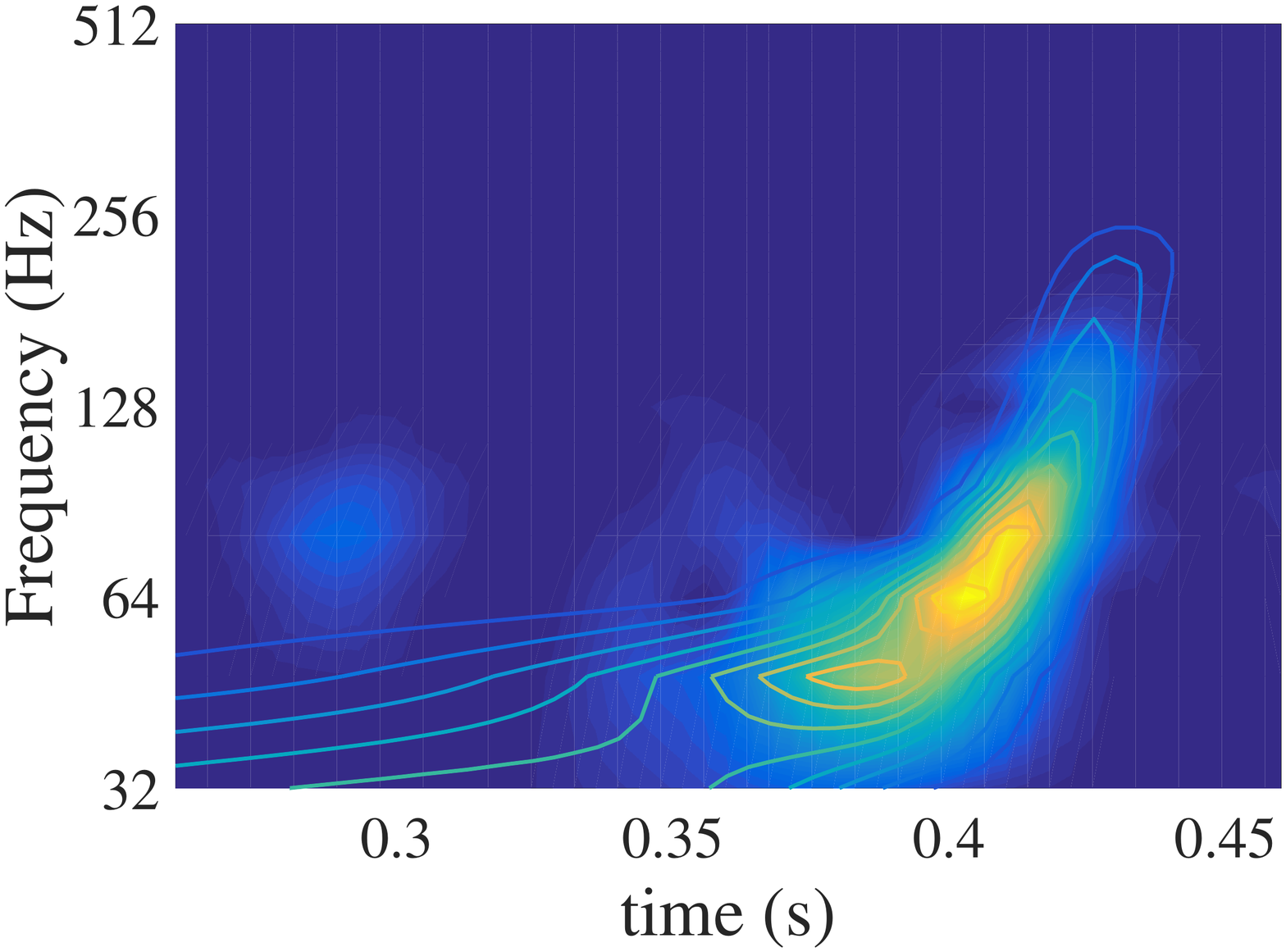}}
    \end{subfigure}
    \hspace{-10mm}
    \begin{subfigure}
    {\includegraphics[width=90mm]{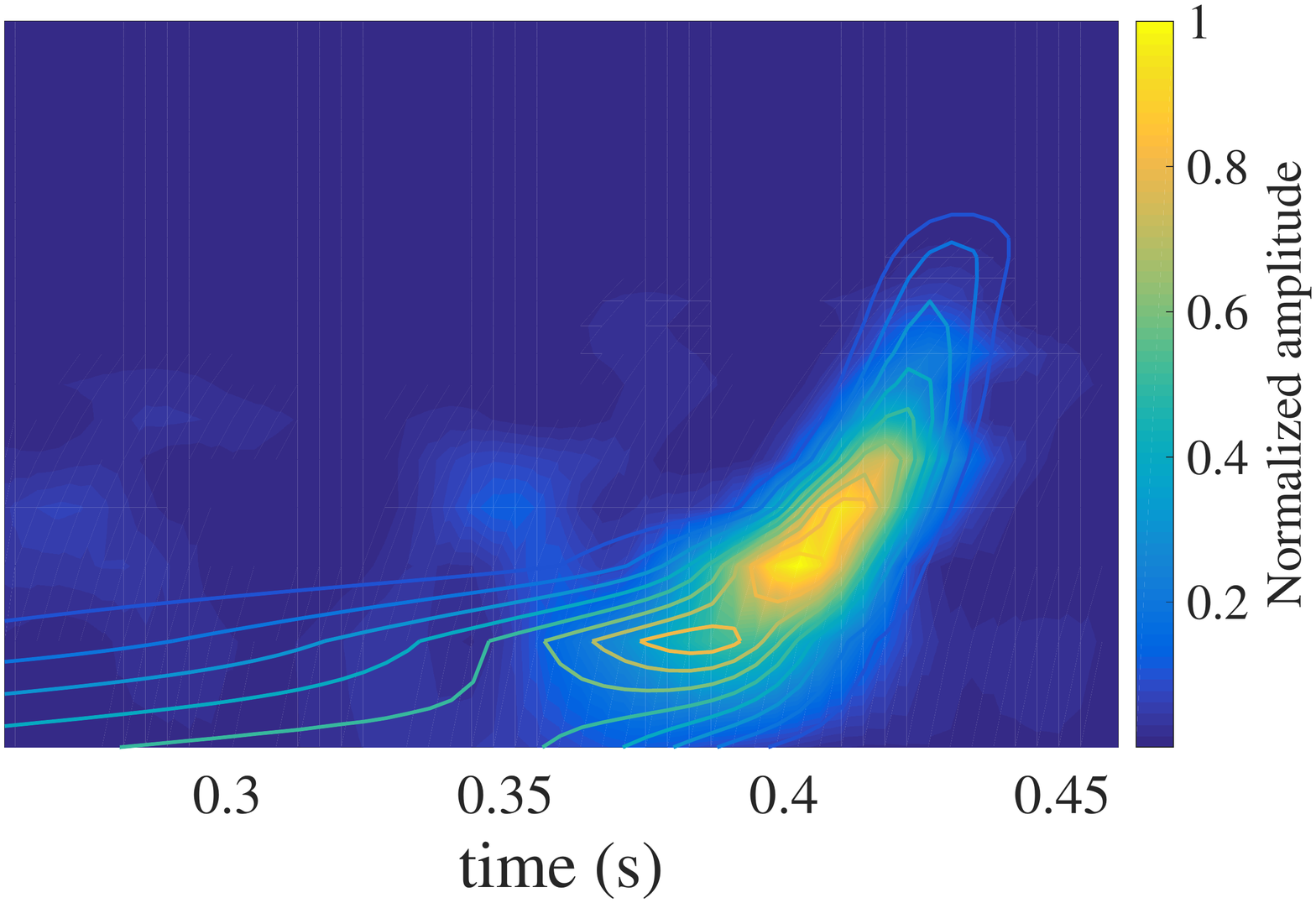}}
    \end{subfigure}\\
    \vspace{-5mm}
    \caption{Spectrograms of our denoised waveforms after
    applying the rROF algorithm to the data from Hanford (left panel) and Livingston (right panel). The superimposed isocontours     correspond to the numerical relativity waveform template.}
    \label{fig:spectrogram}
\end{figure*}

The denoised waveforms for both the Hanford and Livingston LIGO detectors are shown in the top panel of Fig.~\ref{fig:Strain}. 
For both detectors the waveforms look remarkably similar, especially in the last few cycles of the signal.
Comparing these denoised waveforms with the filtered waveforms reported in~\cite{GW150914-prl} we notice that the latter show high-frequency features visible in some of the final cycles before the merger that do not appear in our results. 
By reducing the number of iterations to about 3-5 and decreasing the value of $\mu$, our rROF algorithm does 
not entirely smooth out higher frequencies locally which results in a somewhat closer similitude between both sets 
of waveforms. 

In order to make a fair comparison, we have applied the rROF method to both, the observed data and to the same numerical relativity waveform employed by~\cite{GW150914-prl}. This comparison is shown in the bottom panel of Fig.~\ref{fig:Strain}.
The shape of the denoised waveform agrees with the shape of the processed numerical relativity waveform template of~\cite{abbott-1} for the last 4 cycles and the ringdown part of the signal~\footnote{Note that since we do not reconstruct the waveform but just denoise it, we cannot present the same comparison reported in the middle panel of Fig.~1 in~\cite{GW150914-prl} between the waveform reconstructed through a Bayesian approach and the reconstructed numerical relativity template.}. To have a quantitative measure of the quality of our results, we calculate the Mean Square Error (MSE) between our denoised waveform and the numerical relativity waveform in the time interval shown in Fig.~\ref{fig:Strain}. We also perform the same calculation with the data resulting from applying a whitening process to the original H1 and L1 data and filtering them with a  35-350 Hz bandpass filter. For the Hanford detector our results yield a MSE value of $0.0195$ while for the whitened and filtered data the MSE value is $0.0489$. These low values of the MSE show that both methods remove noise successfully. 

As in Ref.~\cite{GW150914-prl} we have also computed the spectrograms (time-frequency diagrams) of the waveforms for our denoised signals. These are shown in Fig.~\ref{fig:spectrogram} for the signals of both detectors. The superimposed isocontours appearing in both spectrograms correspond to the numerical relativity waveform. These lie on top of the time-frequency diagrams of the denoised signals. The distinctive increase in frequency during the final cycles of the chirp signal is clearly recognizable in both spectrograms. Our results are again similar to the results reported by~\cite{GW150914-prl}. 

\section{Summary}
\label{section:summary}

We have applied a regularized Rudin-Osher-Fatemi total variation algorithm to denoise the transient gravitational 
wave signal GW150914 detected on September 14, 2015 by the two Advanced-LIGO interferometers. 
 Our goal has been to assess if the rROF algorithm applied in the time domain to real gravitational wave data
can successfully remove noise without any a priori information about the signal.
We have recently applied TV-denoising techniques to numerically generated gravitational 
waves embedded in additive Gaussian noise, obtaining satisfactory results irrespective of the signal morphology 
or astrophysical origin~\cite{alex}. The results reported in this paper show that the non-Gaussian, non-stationary noise 
from an actual observation such as the gravitational wave event 
GW150914 can also be successfully removed with TV-denoising methods. The quality of the denoised 
waveform is comparable to that obtained with the Bayesian approach used in the discovery 
paper~\cite{GW150914-prl}. TV-denoising techniques may offer thus an additional viable approach for waveform reconstruction. 
As we already pointed out in~\cite{alex} TV-denoising algorithms should also be
useful to improve the results of other data analysis approaches
such as Bayesian inference or matched filtering when used as
a noise removal initial step that might induce more accurate
results for those other methods. With the potential increase in the wealth of new gravitational wave data in the near future, investigating
the performance of TV-denoising methods such as the iterative rROF algorithm presented in this work, particularly in less favourable SNR conditions than those present in GW150914, deserves further analysis.

\begin{acknowledgments}
We thank the LIGO Scientific Collaboration for releasing the
strain time-series data associated with the gravitational wave event GW150914.
Work supported by the Spanish MINECO (grants AYA2013-40979-P, MTM2014-56218-C2-2-P) and by
 the {\it Generalitat Valenciana} (PROMETEOII-2014-069).
 \end{acknowledgments}
 


\begin{thebibliography}{99}
\bibitem{GW150914-prl}
B.P.~Abbott et al., Phys.\ Rev.\ Lett. {\bf 116}, 061102 (2016).

 \bibitem{campanelli:2006}
 M.~Campanelli, C.~O.~Lousto, P.~ Marronetti, and Y.~Zlochower. Phys.\ Rev. \ Lett. {\bf 96}, 111101 (2006).

\bibitem{SXS}
Signal SXS:BBH:0305, available at URL  \url{http://www.black-holes.org/waveforms.} 

\bibitem{Mroue:2013}
A.H~Mrou\'e et al, Phys.\ Rev. \ Lett. {\bf 111}, 241104 (2013).

\bibitem{abbott-1}
B.~Abbott et al, \url{https://doc.ligo.org/LIGO-P1500218/public/main.}

\bibitem{cornish:2015}
N.J.~Cornish and T.B.~Littenberg,   Class. Quantum Grav. {\bf 32}, 135012 (2015).      

\bibitem{abbott-2}
B.~Abbott et al, \url{https://doc.ligo.org/LIGO-P1500229/public/main.}

\bibitem{alex}
A.~Torres, A.~Marquina, J.A.~Font, and J.M.~Ib\'a\~nez, Phys. Rev. D {\bf  90}, 084029 (2014).

\bibitem{Goldstein:2009}
T.~Goldstein and S.~Osher, J. Imag. Sci. {\bf 2}, 323 (2009).

\bibitem{Rudin:1992}
L.I.~Rudin, S.~Osher and E.~Fatemi, Physica {\bf D}, 60 259-268 (1992).

\bibitem{Bregman:1967}
L.~Bregman ,USSR Computational Mathematics and Mathematical Physics, {\bf 7}, 200-217 (1967).

\bibitem{Osher:2005}
S.~Osher, M.~Burger, D.~Goldfarb, J.~Xu, and W.~Yin, Physica. D {\bf 60}, 259-268 (2005).

\bibitem{GW150914data}
\url{https://losc.ligo.org/events/GW150914/}

\end{thebibliography}
\end{document}